\documentclass[prl,twocolumn,amsmath,amsfonts,amssymb,showpacs]{revtex4}

\usepackage{graphicx}
\usepackage{bm}

\begin{document}
\title{Pumping-Restriction Theorem for Stochastic Networks.}

\author{V. Y. Chernyak}
\affiliation{Department of Chemistry, Wayne State University, 5101 Cass Ave,Detroit, MI 48202}

\affiliation{Theoretical Division, Los Alamos National Laboratory, Los Alamos, NM 87545}

\author{N. A. Sinitsyn}
\affiliation{Center for Nonlinear Studies and Computer, Computational
  and Statistical Sciences Division, Los Alamos National Laboratory,
  Los Alamos, NM 87545 USA}


\begin{abstract}
We formulate an exact result, which we refer to as the pumping restriction theorem (PRT). It imposes strong restrictions on the currents generated by periodic driving in a generic dissipative system with detailed balance.
Our theorem unifies previously known results with the new ones and provides a universal nonperturbative
approach to explore further restrictions on the stochastic pump effect in non-adiabatically driven systems.
\end{abstract}

\date{\today}

\maketitle

{\em Introduction}. The stochastic pump effect (SPE), which is the rectification of classical stochastic currents under
time-periodic excitations, manifests itself in various systems 
\cite{tsong}. For example, it  plays a fundamental role in the theory
of molecular motors \cite{westerhoff-86}.
A theory of the SPE, developed in the limit of adiabatically slow perturbations \cite{astumian-pnas,sinitsyn-07epl},
provided  qualitative insight as well as proposed quantitative approaches for pump-current calculations.

Beyond the adiabatic limit, such a unifying theory is yet to emerge. Although many specific systems have been studied in great detail,
universal results in this domain have been very rare.
The recently discovered Fluctuation Theorems proposed a new approach to the  nonadiabatic regime \cite{jarzynski-review}.
They addressed such questions as work production or probability relations
among single trajectories  \cite{andrieux-07},  but not the magnitude of the SPE.

In a recent work of Rahav, Horowitz and Jarzynski \cite{jarzynski-08},  a universal result for the non-adiabatic regime, called the ``no-pumping'' theorem, was formulated. 
One can parametrize kinetic rates of an arbitrary Markov chain with detailed balance conditions, so that for any pair of sites $i$ and $j$ these rates are written as
$k_{ij} = k e^{E_j-W_{ij}}$, where $E_i$ can be called the depth of a potential well $i$, and $W_{ij}=W_{ji}$ is called the size of the potential barrier $i$-$j$. Energy scale is $k_BT=1$.
The theorem \cite{jarzynski-08} states that to generate rectified currents
during a cyclic process, both well depths and barrier sizes must be varied.

In this Letter, we derive  generic restrictions on the SPE, that include previously found theorems  \cite{astumian-pnas,jarzynski-08} as a special case. We show that there is a wide class of stochastic models on graphs, 
where nonadiabatic but periodic modulation of some parameters leads to a zero time-averaged  flux through any link. We also
predict restrictions on the values of nonzero pump currents when they are allowed.
Such restrictions are sensitive to the topology of the graph representing the system.
Thus, we claim that it is possible to determine whether two given links on a
graph belong to the same loop by merely measuring the presence of the SPE.

\begin{figure}[t]
\centerline{\includegraphics[width=5 cm]{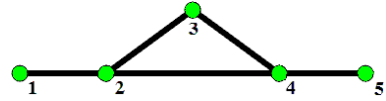}}
\caption{\label{loop15} Five-state graph with one loop $[2342]$. }
\end{figure}

{\em Representative examples.} Consider a particle moving according to Markov chain rules on a graph with a finite number of states and the detailed balance.
First, we note a trivial observation, that periodic variation of parameters on a tree-like graph
 would not produce a net current because any flux that passes through any given link would eventually
return through the same link after driven parameters return to the initial values.
Pump currents are possible only on graphs with loops, such as the one shown in Fig.~\ref{loop15}.
Note, however, that the
previous analysis still applies to some of the links. Namely, the integrated over time fluxes through links 1-2 and 4-5 in Fig.~\ref{loop15} must be zero.

A less trivial problem is whether there are general conditions (beyond the situation discussed in \cite{jarzynski-08}) under which the flux through any link belonging
 to the loop in Fig~\ref{loop15},
such as the link 2-3 can be zero.
For the model in
Fig~\ref{loop15} we claim, that if one varies the rates  related to the links 1-2 and 4-5 cyclically but otherwise arbitrarily, while other rates remain constant and satisfy the detailed balance condition,
then the time integrated current through any link on
the loop will be zero.
To prove it, we split the set of links in Fig.~\ref{loop15} into two
subsets $\{$1-2, 4-5$\} \in X_1$ and $\{ $2-3, 3-4, 2-4$ \} \in X_0$.
Consider the evolution of probabilities $p_i$ on three sites connected by the links in  $X_0$, i.e. the sites with indexes $i=2,3,4$.
Conservation laws require that
\begin{equation}
\begin{array}{l}
\dot{p}_2 = j_{12}(t)-j_{23}(t)-j_{24}(t),\\
\dot{p}_3 = j_{23}(t)-j_{34}(t),\\
\dot{p}_4 = j_{34}(t)+j_{24}(t)-j_{45}(t),\\
\end{array}
\label{eq1}
\end{equation}
where $j_{ij}(t)$ is the current passing through the link $i$-$j$. 
Requiring the final probability distribution to be equal to the initial one, we reach the conditions
\begin{equation}
\int_0^Tdt \dot{p}_i = \int_0^T dt j_{12}(t) = \int_0^T dt j_{45}(t) = 0,
\label{zero}
\end{equation}
where the upper limit of the integration depends on the choice of the driving protocol. In case of a single localized
pulse that drives kinetic rates on links in $X_1$, $T$ must be  formally infinite in order to allow the system to relax to the equilibrium.
In case of a steady periodic driving with the period $\tau$, if the system already reached the steady regime
 with $p_i(t)=p_i(t+\tau)$, one can choose the upper integration limit $T=\tau$.

\begin{figure}[t]
\centerline{\includegraphics[width=7 cm]{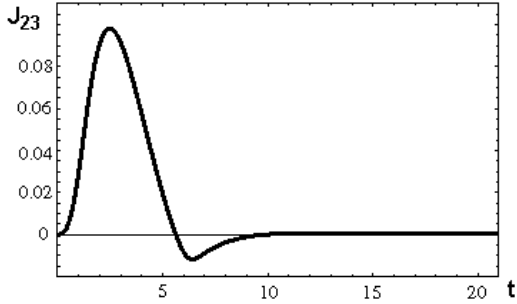}}
\caption{\label{check1} The total flux passed through the link $2$-$3$ by time $t$. Variable parameters are $E_1(t)=E_1+E\sin(\omega t)$,
 and $W_{45}(t)=W_{45} + W \cos(\omega t)$,
for $t \in (0,2\pi/\omega)$, and $E_1(t)=E_1$, $W_{45}(t)=W_{45}+W$ for $t > 2\pi/\omega$, and we distinguish between $E_1(t)$ and $E_1$ {\em etc.}
Choice of constant parameters is $\omega = 1$, $E=1$, $W=2$, $E_1=-1$, $E_2=0.2$, $E_3=-0.05$, $E_4=0.1$, $E_5=-0.25$, $W_{45}=0.3$, $W_{12}=0$, $W_{23}=0.2$, $W_{34}=-0.2$, $W_{24}=0.25$, 
$W_{ij}=W_{ji}$ for any $i$-$j$.}
\end{figure}
The current through any link can be formally written in terms of instantaneous site probabilities  and the
kinetic rates,
$j_{ij}=k_{ji}p_{i} - k_{ij}p_j.$
Denote $\rho_i \equiv  \int_0^T p_i(t)dt$.
Integrating (\ref{eq1}) over time, using conditions (\ref{zero})  one arrives at a set of equations,
\begin{equation}
\begin{array}{l}
-(k_{32}+k_{42})\rho_2  +k_{23}\rho_3+k_{24}\rho_4=0,\\
-(k_{23}+k_{43}) \rho_3 +k_{32}\rho_2+k_{34}\rho_4=0,\\
-(k_{24}+k_{34}) \rho_4 +k_{42}\rho_2+k_{43}\rho_3=0.\\
\end{array}
\label{eq2}
\end{equation}
The set of equations (\ref{eq2}) for $\rho_i$ coincides with the one for probabilities on
 a 3-state Markov chain with the same topology and rates as for a subset $X_0$ at equilibrium state.
 Since all rates on this subset are time-independent and satisfy the detailed balance, the solution of (\ref{eq2}) is
$\rho_i=Ce^{-E_i}$, $ i=2,3,4$,
where $C$ is a constant that depends on the details of the driving protocol, but is equal for all three states on the loop.
The total flux passed through e.g. the link 2-3 then reads
\begin{equation}
J_{23}(T) \equiv \int_0^T j_{23}(t) dt = C (k_{32}e^{-E_2}-k_{23} e^{-E_3}) = 0.
\label{sol2}
\end{equation}
This concludes our proof that varying the  kinetic rates outside the loop only, does not lead to a net time averaged flux through any link on the graph  in Fig.~\ref{loop15}.

{\em Numerical check}. To test our predictions we performed numerical simulations for the graph shown in Fig.~\ref{loop15}. In all examples we assume the kinetic  rates to
satisfy the detailed balance condition with the parametrization: $k_{ij}=k\exp([E_j-W_{ij}]/k_BT)$, \cite{jarzynski-08}. We chose the energy scale so that
$k=1$, $k_BT=1$. 
 Fig.~\ref{check1} shows the
flux passed through the link 2-3 on the loop, when only parameters $E_1$ and $W_{45}$ outside the loop are changing with time. After completion of the external driving, the system
relaxes with a zero net current through the link 2-3, as we predicted.
\begin{figure}[t]
\centerline{\includegraphics[width=7 cm]{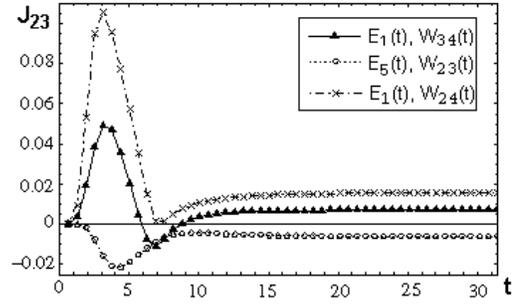}}
\caption{\label{check2} The total flux passed through the link 2-3 by time $t$. $W_{ij}$ is varied along one of the loop links. Variable parameters, indicated
on a graph legend, change with time according to $E_i(t)=E_i+E\sin(\omega t)$, and $W_{ij}(t)=W_{ij} + W \cos(\omega t)$,
for $t \in (0,2\pi/\omega)$, and $E_i(t)=E_i$, $W_{ij}(t)=W_{ij}+W$ for $t > 2\pi/\omega$.
Constant parameters are as in Fig.~\ref{check1}.}
\end{figure}

\begin{figure}[t]
\centerline{\includegraphics[width=7 cm]{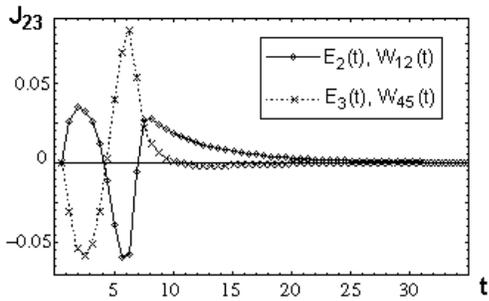}}
\caption{\label{check3} The total flux passed through the link 2-3 by time $t$, when only $E_i$ were allowed to vary on one of the loop sites and $W_{ij}$ is varied along external links. Variable parameters, indicated
on a graph legend, change with time according to $E_i(t)=E_i+E\sin(\omega t)$, and $W_{ij}(t)=W_{ij} + W \cos(\omega t)$,
for $t \in (0,2\pi/\omega)$, and $E_i(t)=E_i$, $W_{ij}(t)=W_{ij}+W$ for $t > 2\pi/\omega$.
Constant parameters are as in Fig.~\ref{check1}.}
\end{figure}

Our numerical results show that if
barriers are varied on the loop but potential depths $E_j$ are varied only on external sites then nonzero pump current through the loop  links appears, as shown in Fig.~\ref{check2}.
However, the opposite is not true. Namely, if one varies well depths on the loop sites but barriers
are varied along external links, numerically we always found no overall pump flux, as we show in
Fig~\ref{check3}. The latter result is also easy to understand by introducing the quantities
$f_i=\int_0^T dt [e^{E_i(t)}p_i(t)]$, $i=2,3,4$. The same analysis as before leads to the
equilibrium master equation for $f_i$ with equal forward/backward rates connecting any pair of
sites, and a constant solution, ensuring the zero time-averaged current through any link on the
loop. The previously found theorem \cite{jarzynski-08} can be proved for any graph by the same steps and
change of variables.  We are now in a position to frame our result in a form of a general and
rigorous mathematical statement.

{\em Pumping-Restriction Theorem (PRT).}
 For a graph $X$ we denote the vector spaces of time averaged populations
${\bm \rho}\in C_{0}(X)$ and time-averaged currents ${\bm J}\in C_{1}(X)$,
 so that ${\bm \rho}=\{\rho_{a}\}$, ${\bm
J}=\{J_{ab}\}$ with $J_{ba}=-J_{ab}$ and $J_{ab}=0$ when the nodes $a$ and $b$ are not connected by
an edge. We further introduce the boundary operator $\partial:C_{1}(X)\to C_{0}(X)$ by
$(\partial{\bm J})_{a}=\sum_{b}J_{ba}$. Let $H_{1}(X)$ be the     
subspace of physical time-averaged currents, i.e. satisfying the continuity condition, and
$H_{0}(X)$  represent the space of 
populations that would be constant within the connected components of $X$. Note that the conjugate
operator $\partial^{\dagger}:C_{0}(X)\to C_{1}(X)$ has a form
$(\partial^{\dagger}{\bm\rho})_{ab}=\rho_{a}-\rho_{b}$. In detailed balance the rates are given by
$k_{ab}=g_{ab}e^{E_{b}}$, where $g_{ab}=g_{ba}=ke^{-W_{ab}}$ is referred to as a metric on $X$. The Euler
theorem claims \cite{comment}
\begin{eqnarray}
\label{Eeuler-theorem}  {\dim}C_{1}-{\dim}C_{0}={\dim}H_{1}-{\dim}H_{0}.
\end{eqnarray}
Consider a partition $X=X_{0}\cup X_{1}$, where $X_{0}$ consists of edges (and adjacent nodes)
where the rates are given by $k_{ab}(t)=g_{ab}e^{E_{b}(t)}$, whereas $X_{1}$ represents the rest of
the edges with arbitrary rates $k_{ab}(t)$. Note that $X_{0}\cap X_{1}$ does not have any edges,
whereas its nodes provide the currents between $X_{0}$ and $X_{1}$. The PRT claims that 

(i) for the described periodic driving the generated pumped current is
restricted  to a vector subspace $V$ such that ${\bm J}\in V\subset H_{1}(X)$, with the dimension
\begin{eqnarray}
\label{dim-V} {\dim}V={\dim}H_{1}(X)-{\dim}H_{1}(X_{0}) ={\dim}C_{1}(X_{1})- \nonumber\\
-{\dim}H_{0}(X_{0})-{\dim}C_{0}(X)+{\dim}C_{0}(X_{0})+1.
\end{eqnarray}
 (ii) given a set of links with driven barriers, the embedding $V\subset H_{1}(X)$ is totally determined by the (time-independent) metric
$g_{ab}$ on $X_{0}$. 

The statement (ii) implies that if there are constraining equations that determine relations among possible pumped currents 
through different links on $X_0$, then
coefficients in these equations will depend only on the 
metric on $X_0$.
Note also that the choice
$X_{0}=X$ reproduces the second no-pumping theorem of \cite{jarzynski-08}. The PRT, however, claims
more: starting with the no-pumping situation with at least one of $E_i$ driven, and also driving the barriers at a certain
number $n$ of links, the number of independent generated currents may not exceed $n$, i.e. $ {\dim}V \leq {\dim}C_{1}(X_{1})$, 
that follows from second equality in (\ref{dim-V}) and obvious inequalities ${\dim}H_0(X_0) \geq 1$, and $[-{\dim}C_{0}(X)+{\dim}C_{0}(X_{0})] \leq 0$. 
Each of the
driven links can be viewed as either responsible for an independent cycle or for connecting two
disconnected parts.  
Therefore, the PRT can be interpreted as the claim of the number of
independent generated currents to be equal to the maximum number of driven barriers, which removal does not split the remaining graph into disjoined 
components.

The proof of PRT is based on the Master Equation and the expression
for the current
\begin{equation}
\label{ME-j} \dot{\bm p}(t)=\partial{\bm j}(t),  \quad {\bm j}(t)=\hat{g}\partial^{\dagger}
e^{\hat{E}(t)}{\bm p}(t),
\end{equation}
where the second equality is valid on $X_{0}$. Representing the current ${\bm j}(t)={\bm
j}_{0}(t)+{\bm j}_{1}(t)$ as the sum of the $X_{0}$ and $X_{1}$ components and averaging
(integrating) Eq.~(\ref{ME-j}) over time we obtain for the time-integrated quantities
\begin{equation}
\label{ME-J-averaged} \partial\hat{g}\partial^{\dagger}{\bm f}=-{\bm\zeta} , \;\;\;
{\bm\zeta}=\partial{\bm J}_{1}|_{X_{0}\cap X_{1}}, \;\;\; {\bm
J}_{0}=\hat{g}\partial^{\dagger}{\bm f},
\end{equation}
where ${\bm f}=\int_{0}^{T}e^{\hat{E}(t)}{\bm p}(t)$, ${\bf J}_{1/0}= \int_{0}^{T}{\bm j}_{1/0}(t)$
and ${\bm\zeta}\in H_{0}(X_{0}\cap X_{1})$ describes the flux passed between $X_{0}$ and $X_{1}$. Since
the operator $\partial\hat{g}\partial^{\dagger}$ can be viewed as the discrete Laplacian in $X_{0}$
associated with the metric $g$, the first two equations in (\ref{ME-J-averaged}) have a solution if
and only if the total flux entering any connected component of $X_{0}$ and $X_{1}$ is zero. Let
$V_{0}\subset H_{0}(X_{0}\cap X_{1})$ is the vector subspace of ${\bm\zeta}$ that satisfies these
conditions. The solution of the first equation in (\ref{ME-J-averaged}) is unique up to an additive
constant distribution, which does not affect the value of ${\bm J}_{0}$, i.e., the latter is
uniquely determined by ${\bm\zeta}\in V_{0}$ and metric on $X_0$. On the other hand, given ${\bm\zeta}\in V_{0}$, the
component ${\bm J}_{1}$ is defined up to a current that circulates completely within $X_{1}$ and is
represented by an element of $H_{1}(X_{1})$. Thus we proved that the metric on $X_0$ and topology of $X_1$ determine the restrictions on ${\bf J}$, which implies the statement (ii) of PRT. In addition our analysis implies 
\begin{eqnarray}
\label{dim-V-implicit} {\rm dim}V={\rm dim}V_{0}+{\rm dim}H_{1}(X_{1}).
\end{eqnarray}
\begin{figure}[t]
\centerline{\includegraphics[width=5 cm]{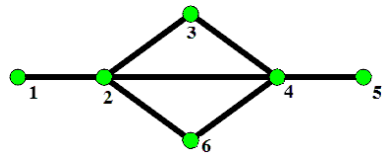}}
\caption{\label{markov6} A six state Markov chain. }
\end{figure}
 To derive the explicit expression (\ref{dim-V}), which completes the proof we combine
Eqs.~(\ref{dim-V-implicit}) with the identity
\begin{eqnarray}
\label{dim-MV-short-sequence} {\rm dim}H_{1}(X)={\rm dim}H_{1}(X_{0})+{\rm dim}H_{1}(X_{1})+{\rm
dim}V_{0},
\end{eqnarray}
which results in the first equality in (\ref{dim-V}). The second equality follows from (\ref{Eeuler-theorem}) and that ${\dim}H_0(X)=1$. Note that the identity (\ref{dim-MV-short-sequence}) has a very simple physical meaning:
For any physical current ${\bm J}\in H_{1}(X)$ on our graph we can identify the current ${\bm
J}^{(01)}\in V_{0}$ that flows from $X_{0}$ to $X_{1}$, and once the exchange current ${\bm
J^{(01)}}$ is identified, the complete current ${\bm J}$ is defined up to the currents ${\bm
J}^{(0)}\in H_{1}(X_{0})$ and ${\bm J}^{(1)}\in H_{1}(X_{1})$ that circulate strictly within
$X_{0}$ and $X_{1}$, respectively.

We further illustrate the PRT using a graph in Fig.~\ref{markov6}, with ${\rm dim}C_1(X)=7$ (the 
 number of links), ${\rm dim}C_0(X)=6$  (the number of sites), and two independent
loops, e.g., $[2342]$ and $[2462]$ that form a basis in $H_{1}(X)$ with ${\rm dim}H_{1}(X)=2$. If
only one barrier is driven we can generate not more than a $1D$ subspace of currents, i.e. ${\rm
dim}V\le 1$. If the barrier at links 1-2 or 4-5 are driven, we are at the no-pump situation, since
${\rm dim}V=0$  (the subgraph $X_{0}$ obtained upon elimination of the link 1-2, or 4-5, has two loops, hence ${\dim}H_1(X_0)=2$). Driving the barrier at any other single link yields
${\rm dim}V=1$, since the graph $X_{1}$ in this case is connected. When a pair of barriers is
driven we have ${\rm dim V}=0$ for $\{$1-2, 4-5$\}$; ${\rm dim}V=1$ for $\{$2-3, 3-4$\}$, $\{$2-6,
6-4$\}$, and when one of 1-2, 4-5 and one of the rest are driven. In all other cases we have ${\rm
dim}V=2$ (all currents may be generated). Consider in more detail the $\{$2-3, 3-4$\}$ driving. The
subgraph $X_{1}$ includes these two links and the vertices $\{2,3,4\}$, $X_{0}$ includes the rest
of the links and the vertices $\{1,2,6,4,5\}$, whereas $X_{0}\cap X_{1}$ has the vertices $\{2,4\}$
and no links. The exchange goes through two nodes $\{2,4\}$, the exchange currents satisfy 
 relations $J_{23}+J_{43}=0$ that yields ${\rm   
dim}V_{0}=2-1=1$, i.e. while allowed elements of $X_{0}\cap X_{1}$ have the form $\{\rho_2,\rho_4\}$, elements in $V_0$ are restricted to be of the form $\{J,-J\}$. The subgraph $X_{1}$ has no loops, and, therefore, no internal currents, which
yields ${\rm dim V}=1$, which agrees with the PRT.


We also note that similar arguments lead to continuous (Langevin dynamics) counterparts of the
no-pumping theorems of \cite{jarzynski-08}.
Let $g^{ij}({\bm x})$ be a metric in a compact oriented manifold $M$ that describes bath-induced fluctuations/dissipation. If parametrized as 
 $g^{ik}({\bm x})=h^{ik}({\bm x})e^{V({\bm x},t)}$, where $h$ is time-independent and $V$ is a periodic in 
time potential, then $J^{j}({\bm x})=0$.

{\em Conclusion.}
Many emerging mesoscopic devices, including molecular motors \cite{astumian-pnas} and nanoscale
electronic circuits \cite{usuki-lz} can be modeled as discrete connected entities with
stochastic transitions among different states.  The control
over such new devices is impossible without deep understanding of non-adiabatic strongly driven
regimes of their operation. Our result is a step toward such a theory. The PRT, presented here, determines the restrictions in the space of
pumped current values on a graph. No-pumping conditions follow as
its special consequences. The restrictions on the
pump current space suggest, for example, that an application of a periodic stimulus can be used to
induce a localized rectified effect without perturbing the whole circuit on average even if all its
components are connected. It should be useful to explore how the PRT is modified by quantum effects.
Violation of the ``no-pumping'' conditions due to quantum corrections can be employed to detect the quantum pump effect \cite{moskalets-review} by
cooling the electronic circuit in the ``no-pumping'' regime down to the quantum domain.

Another possible application for this work is the reconstruction of the stochastic network topologies, e.g.
in biochemical reactions. Standard measurement techniques, such as those based on the linear
response, appear insufficient. One would expect to find a nonzero signal anywhere on  an ergodic
Markov chain in response to a time-dependent current inducing perturbation. However, we showed that
measuring rectified current in response to external periodic stimulus can help one to identify
whether or not two given links belong to the same loop. One of the perturbed links must
belong to a common loop with the measured one in order to observe the SPE. Even without a
quantitative understanding of the data, detecting only
the presence of the SPE can be sufficient to deduce the topological structure of the network completing only a small number of measurements.

\begin{acknowledgments}
We thank the authors of Ref. \cite{jarzynski-08} for sharing their results prior to publication, and
also B. Munsky and I. Nemenman for useful discussions. This material is based upon work supported
by the National Science Foundation under CHE-0808910 and in part by DOE under Contract No.\
  DE-AC52-06NA25396.
\end{acknowledgments}


\begin{thebibliography}{100}
\bibitem{comment} It follows from identities $H_{1}(X)={\rm ker}\partial$, $H_{0}(X) = C_{0}(X)/{\rm im}\partial$, 
$C_1(X) \cong  {\rm ker \partial} \oplus C_1/{\rm ker} \partial $ and $C_0(X) \cong  {\rm im \partial}\, \oplus \, C_0/{\rm im} \partial $. Taking dimensions in both parts of 
all equations and using $C_1/{\rm ker} \partial \cong {\rm im \partial}$, we 
arrive at  (\ref{Eeuler-theorem}).
\end{thebibliography}
\end{document}